\documentclass[prl,aps]{revtex4}
\usepackage{psfrag}
\usepackage{graphicx}
\usepackage{dcolumn}
\usepackage{color}
\usepackage{latexsym,amsfonts}
\usepackage{bm}
\usepackage{amssymb}
\begin{document}

\noindent {\bf Reply by A.N.Ivanov and P. Kienle}: Nowadays it is well
established experimentally that neutrinos $\nu_{\alpha}$ with lepton
flavours $\alpha = e, \mu$ and $\tau$ are superpositions
$|\nu_{\alpha}\rangle = \sum_jU^*_{\alpha j}|\nu_j\rangle$ of massive
neutrino mass--eigenstates $|\nu_j\rangle$ with masses $m_j$, where
$U^*_{\alpha j}$ are elements of the $3\times 3$ unitary mixing matrix
$U$, defined by mixing angles $\theta_{ij}$ \cite{PDG08}. The wave
functions $|\nu_{\alpha}\rangle$ and $|\nu_j\rangle$ are orthogonal
and used for the description of neutrino oscillations $\nu_{\alpha}
\longleftrightarrow \nu_{\beta}$ with frequencies $\omega_{ij} =
\Delta m^2_{ij}/2E$, where $E$ is the neutrino energy and $\Delta
m^2_{ij} = m^2_i - m^2_j$ \cite{PDG08}. In K--shell electron capture
$(EC)$ decays of the H--like heavy ions $m \to d + \nu_e$, where $m$
and $d$ are mother and daughter ions in their ground states
\cite{GSI,Ivanov}, one deals with an emission of electron neutrinos
$|\nu_e\rangle = \sum_jU^*_{e j}|\nu_j\rangle$. Thus, the $EC$--decay
rates of the H--like heavy ions are defined by the decay channels $m
\to d_j + \nu_j$, where the final states are described by the
orthogonal wave functions $\langle \nu_id_i|d_j\nu_j\rangle = 0$ for
$i\neq j$. The states of the daughter ions $d_j$ differ in 3--momenta
$\vec{q}_j$ and energies $E_d(\vec{q}_j)$. The massive neutrinos
$\nu_j$ are produced with 3--momenta $\vec{k}_j= - \vec{q}_j$ and
energies $E_j(\vec{k}_j)$, caused by conservation of energy and
momentum in the decay channels $m \to d_j + \nu_j$. Since massive
neutrinos $\nu_j$ are not detected they appear in the asymptotic
states with 3--momenta $\vec{k}_j$, energies $E_j(\vec{k}_j)$ and
energy differences $\omega_{ij} = \Delta m^2_{ij}/2M_m$, where $M_m$ is
the mass of the mother ion $m$ \cite{Ivanov}.  In the GSI experiments
\cite{GSI,Ivanov} the $EC$--decay channels $m \to d_j + \nu_j$ are
measured by detecting the daughter ions $d_j$. If the daughter ions
would be detected in the asymptotic states with 3--momenta $\vec{q}_j$
and energies $E_d(\vec{q}_j)$, the probability per unit time of the
$EC$--decay $m \to d + \nu_e$ is equal to
$$
P(m\to d\,\nu_e)(t) =
\sum_j |U_{e j}|^2P(m\to d_j\,\nu_j)(t) = \sum_j |U_{e
j}|^2\frac{d}{dt}|A(m\to d_j\nu_j)(t)|^2, \eqno(1)
$$ where $A(m\to d_j\nu_j)(t)$ is the amplitude of the decay channel
$m \to d_j + \nu_j$. However, this is not the case in the GSI
experiments, where the time differential detection of the daughter
ions with a time resolution $\tau_d$ leads to indistinguishability of
daughter ions in the decay channels $m \to d_j + \nu_j$.  As a result
the daughter ions $d_j$ are measured in the asymptotic state $d$ with
a 3--momentum $\vec{q}$ and an energy $E_d(\vec{q}\,)$ such that
$\vec{q} \simeq \vec{q}_j$ and $E_d(\vec{q}\,) \simeq E_d(\vec{q}_j)$
\cite{Ivanov}. This does not violate the orthogonality of the wave
functions in the final state $\langle \nu_id|d\nu_j\rangle = 0$ for
$i\neq j$. The energy and momentum uncertainties $\delta E_d$ and
$|\delta \vec{q}_{\,d}|$, respectively, induced by the time
differential detection of the daughter ions, provide the overlap of
the wave functions of the daughter ions if $\delta E_d \gg
|\omega_{ij}|$ and $|\delta \vec{q}_{\,d}| \gg |\vec{q}_i -
\vec{q}_j|=|\vec{k}_i - \vec{k}_j|$, where $\omega_{ij}$ present also
the differences of the recoil energies of the daughter ions. The time
differential detection of the daughter ions $d_j$ in the asymptotic
state with the 3--momentum $\vec{q}$ and an energy $E_d(\vec{q}\,)$
results in a smearing of momenta and energies in the decay channels $m
\to d_j + \nu_j$ around $\vec{q}+ \vec{k}_j \simeq 0$ and
$E_d(\vec{q}\,) + E_j(\vec{k}_j) \simeq M_m$. This is the origin of
the non--vanishing interference terms in the probability per unit time
of the $EC$--decay $m \to d + \nu_e$ \cite{Ivanov}
$$ P(m\to d\,\nu_e)(t) = \sum_j |U_{e j}|^2P(m\to d\,\nu_j)(t) + 2
\sum_{i > j} \frac{d}{dt}{\rm Re}[U^*_{e i}U_{e j}A^*(m\to
d\,\nu_i)(t)A(m\to d\,\nu_j)(t)],\eqno(2)
$$ where in comparison with Eq.(1) the second sum is caused by the
interference terms. Since uncertainties $\delta E_d$ and $|\delta
\vec{q}_{\,d}|$ are rather small \cite{Ivanov} and $|\vec{k}_j| =
|\vec{q}_j| \simeq |\vec{q}\,| \simeq Q_{EC}$ and $Q_{EC} \gg m_j$,
where $Q_{EC}$ is the $Q$--value of the $EC$--decay $m \to d + \nu_e$,
one can set the neutrino masses zero everywhere except for energy
differences $\omega_{ij}$ and mixing angles $\theta_{ij}$ in the
interference terms \cite{Ivanov}, and neutrino 3-momenta equal
$\vec{k}_j = \vec{k}$. As a result the $EC$--decay rate is given by
\cite{Ivanov}
$$ \lambda_{EC}(t) = \frac{1}{2M_m}\int P(m\to
d\,\nu_e)(t)\,\frac{d^3q}{(2\pi)^3 2 E_d}\frac{d^3k}{(2\pi)^3 2
E_{\nu_e}} = \lambda_{EC}\Big(1 + 2\sum_{i > j}{\rm Re}[U^*_{e i}U_{e
j}] \cos(\omega_{ij}t)\Big). \eqno(3)
$$ Thus, the asymptotic orthogonality of the final state wave
functions does not influence the observation of the time modulation of
the $EC$--decays, observed in the GSI experiments with a time
resolution $\tau_d \ll T_{ij}$ much shorter than the modulation
periods $T_{ij} = 2\pi/\omega_{ij}$ \cite{Ivanov}. This is unlike the
assertion by Flambaum \cite{Flambaum}. For $\tau_d \gg T_{ij}$ the
daughter ions $d_j$ become distinguishable with 3--momenta $\vec{q}_j$
and energies $E_d(\vec{q}_j)$ the time modulation vanishes (see
Eq.(1)).  For the theoretical description of the GSI data \cite{GSI},
accounting for the procedure for the detection of the daughter ions,
one can use time--dependent perturbation theory and wave packets for
the wave functions of the daughter ions, related to the density matrix
description of unisolated quantum systems \cite{Landau}. For technical
details we refer to \cite{Ivanov}.

\end{document}